\documentclass[aps,prd,showpacs,twocolumn]{revtex4}

\usepackage{amssymb}
\usepackage{amsfonts}
\usepackage{amsmath}
\usepackage{graphicx}
\usepackage{color}
\usepackage[dvips]{epsfig}
\usepackage[dvips]{graphicx}
\usepackage{float}

\begin{document}

\title{Effects of a \emph{CPT}-even and Lorentz-violating nonminimal
coupling on the electron-positron scattering}
\author{R. Casana$^{a}$, M. M. Ferreira Jr$^{a}$, R.V. Maluf$^{b}$, and
F.E.P. dos Santos$^{a}$\thanks{%
e-mails: rodolfo.casana@gmail.com, manojr07@ibest.com.br,}}
\affiliation{$^{a}${\small {Universidade Federal do Maranh\~{a}o (UFMA), Departamento de F%
\'{\i}sica, Campus Universit\'{a}rio do Bacanga, S\~{a}o Luiz, MA,
65085-580, Brazil}}}
\affiliation{$^{b}${\small {Instituto de F\'{\i}sica, Universidade de S\~{a}o Paulo
(USP), Caixa Postal 66318, S\~{a}o Paulo, SP, 05315-970 Brazil}}}

\begin{abstract}
We propose a new \emph{CPT}-even and Lorentz-violating nonminimal coupling
between fermions and Abelian gauge fields involving the \emph{CPT}-even
tensor $\left( K_{F}\right) _{\mu\nu\alpha\beta}$ of the standard model
extension. We thus investigate its effects on the cross section of the
electron-positron scattering by analyzing the process $e^{+}+e^{-}%
\rightarrow\mu^{+}+\mu^{-}$. Such a study was performed for the parity-odd
and parity-even nonbirefringent components of the Lorentz-violating $\left(
K_{F}\right) _{\mu\nu\alpha\beta}$ tensor. Finally, by using experimental
data available in the literature, we have imposed upper bounds as tight as $%
10^{-12}(\mbox{eV})^{-1}$ on the magnitude of the \emph{CPT}-even and
Lorentz-violating parameters while nonminimally coupled.
\end{abstract}

\maketitle

\section{Introduction}

The standard model extension (SME) is a large theoretical framework that
includes terms of Lorentz and \emph{CPT} violation in the structure of the
usual standard model \cite{Colladay}. This model was proposed after the
verification about the possibility of having spontaneous violation of
Lorentz symmetry in the context of string theories \cite{Samuel}. The
Lorentz-violating (LV) terms are generated as vacuum expectation values of
tensor quantities, keeping the coordinate invariance of the extended theory~%
\cite{Lehnert}. This model has been scrutinized in many respects in the
latest years, with studies embracing the fermion and gauge sectors, and
gravitation extension \cite{Gravity}. The fermion sector \cite{Fermion} was
much examined, mainly in connection with \emph{CPT}-violating tests to
impose some upper bounds on the magnitude of the LV terms \cite{CPT},
dealing with other interesting aspects as well \cite{Fermion2}. The Abelian
gauge sector of the SME is composed of a \emph{CPT}-odd \cite{Jackiw} and a
\emph{CPT}-even sector, both intensively investigated in the latest years
\cite{Adam,Cherenkov1,Photons1,KM1,Risse,Cherenkov2,Kostelec}.

Besides the investigations undertaken into the structure of the SME, some
other works were proposed to examine Lorentz-violating developments out of
this broad framework. Some of them involve nonminimal coupling terms that
modify the vertex interaction between fermions and photons. \emph{CPT}-odd
nonminimal couplings as $gv_{\mu}\tilde{F}^{\mu\nu}$and $g\gamma_{5}b_{\mu}%
\tilde{F}^{\mu\nu}$ were considered some time ago in the context of the
Dirac equation, with interesting consequences in the nonrelativistic limit,
involving topological phases \cite{NM1,NM2}, corrections on the hydrogen
spectrum \cite{NM3}. Such nonminimal coupling has been reassessed in
connection with its implications on the Aharonov-Bohm-Casher problem \cite%
{NMabc}, the Bhabha cross section \cite{NMmaluf}, and other respects \cite%
{NMothers}. Recently, other types of nonminimal coupling, defined in the
context of the Dirac equation, have been proposed for investigating the
generation of topological and geometrical phases \cite{NMbakke}.

Theoretical studies about cross section evaluation in the presence of
Lorentz-violating terms were accomplished by some authors \cite{Colladay2},
searching to elucidate the route for evaluating the cross section for a
general scattering. Very recently, some authors performed a study on the
Bhabha scattering \cite{NMmaluf}, determining the effects induced by the
nonminimal \emph{CPT}-odd coupling on the Bhabha cross section. The results
were compared with some available data concerning this scattering \cite%
{Derrick} and used to impose the upper bound $\left\vert gv_{\mu}\right\vert
\leq10^{-12}\left( \mbox{eV}\right) ^{-1}.$

In this work, we reassess a well-known quantum electrodynamics process, the $%
e^{+}+e^{-}\,\rightarrow\mu^{+}+\,\mu^{-}$ scattering, in the presence of a
new Lorentz-violating \emph{CPT}-even nonminimal coupling involving the
fermion and gauge sectors. First, we calculate the scattering amplitude,
considering new Feynman diagrams due to the emergence of a new vertex in the
theory. In order to evaluate the total cross-section, we first calculate the
unpolarized squared amplitude, using the Casimir trick. We specialize our
evaluations for the parity-odd and parity-even subsectors of the \emph{CPT}%
-even gauge sector. At the end, following the approach of Refs. \cite%
{NMmaluf} and \cite{Derrick}, we compare the cross section results with the
experimental data, finding an upper limit for the magnitude for the new
nonminimal coupling as tight as $\left\vert \lambda\left( K_{F}\right)
\right\vert \leq10^{-12}\left( \mbox{eV}\right) ^{-1}.$

\section{The theoretical model}

We are interested in analyzing some aspects of a modified quantum
electrodynamics, whose fermion sector is governed by the generalized Dirac
equation,
\begin{equation}
\left(i\gamma^{\mu}D_{\mu}-m\right)\Psi=0,  \label{Dirac1}
\end{equation}
in which the usual covariant derivative is supplemented by a nonminimal
\emph{CPT}-even coupling term, that is,
\begin{equation}
D_{\mu}=\partial_{\mu}+ieA_{\mu}+\frac{\lambda}{2}\left( K_{F}\right)
_{\mu\nu\alpha\beta}\gamma^{\nu}F^{\alpha\beta},  \label{covader}
\end{equation}
where $\left( K_{F}\right) _{\mu\nu\alpha\beta}$ is the tensor that embraces
the 19 LV terms belonging to the \emph{CPT}-even gauge sector of the SME.
This tensor possesses the same symmetries of the Riemann tensor: $\left(
K_{F}\right) _{\alpha\nu\rho\varphi}=-\left( K_{F}\right) _{\nu\alpha
\rho\varphi},~\left( K_{F}\right) _{\alpha\nu\rho\varphi}=-\left(
K_{F}\right) _{\alpha\nu\varphi\rho},~\left( K_{F}\right) _{\alpha\nu
\rho\varphi}=\left( K_{F}\right) _{\rho\varphi\alpha\nu}$ and a double null
trace, $\left( K_{F}\right) ^{\alpha\beta}{}_{\alpha\beta}=0,$ implying 19
components. Using the symmetries of the tensor $\left( K_{F}\right) _{\mu
\nu\alpha\beta}$ in the Dirac (\ref{Dirac1}) equation, one obtains
\begin{equation}
\left[ i\gamma^{\mu}\partial_{\mu}-e\gamma^{\mu}A_{\mu}+\frac{\lambda}{2}%
\left( K_{F}\right) _{\mu\nu\alpha\beta}\sigma^{\mu\nu}F^{\alpha\beta }-m%
\right] \Psi=0,  \label{DiracM1}
\end{equation}
with
\begin{equation}
\sigma^{\mu\nu}=\frac{i}{2}(\gamma^{\mu}\gamma^{\nu}-\gamma^{\nu}\gamma^{\mu
})=\frac{i}{2}[\gamma^{\mu},\gamma^{\nu}],  \label{OP1}
\end{equation}
whose components, $\sigma^{0i}$ and $\sigma^{ij},$ are
\begin{equation}
\sigma^{0i}=i\left(
\begin{array}{cc}
0 & \sigma^{i} \\
\sigma^{i} & 0%
\end{array}
\right) ,\text{ \ \ }\sigma^{ij}=-\left(
\begin{array}{cc}
\epsilon_{ijk}\sigma^{k} & 0 \\
0 & \epsilon_{ijk}\sigma^{k}%
\end{array}
\right) .
\end{equation}
This new coupling, represented by $(\lambda K_{F})_{\mu\nu\alpha\beta}$, has
mass dimension $[\lambda K_{F}]=-1$, which leads to a nonrenormalizable
theory at power counting. This respect does not pose a problem for this
investigation, once we are interested in analyzing the tree-level scattering
process.

We now present the Lagrangian of the modified QED,
\begin{equation}
\mathcal{L}_{\text{mod}QED}=\mathcal{L}_{QED}+\mathcal{L}_{I}^{new}
\label{L1}
\end{equation}%
where $\mathcal{L}_{QED}$ is the usual Lagrangian density of QED in the
Lorenz gauge,
\begin{equation}
\mathrm{{\mathcal{L}}}_{QED}=\bar{\psi}(i{\rlap{\hbox{$\mskip 1 mu /$}}}%
\partial -e{\rlap{\hbox{$\mskip 1 mu /$}}}A-m)\psi -\frac{1}{4}F_{\mu \nu
}F^{\mu \nu }-\frac{1}{2\xi }(\partial _{\mu }A^{\mu })^{2},
\end{equation}%
and $\mathrm{{\mathcal{L}}}_{I}^{new}~$represents the new interaction
produced by the nonminimal coupling, to be regarded
\begin{equation}
{{\mathcal{L}}}_{I}^{new}=\frac{\lambda }{2}(K_{F})_{\mu \nu \alpha \beta }%
\bar{\psi}\sigma ^{\mu \nu }\psi F^{\alpha \beta }.
\end{equation}%
In the next steps we will consider the Feynman gauge, $\xi =1$. The theory
represented by Lagrangian (\ref{L1}) has, besides the usual vertex, $\bullet
\rightarrow -ie\gamma ^{\mu }$, an additional LV vertex, represented as
\begin{equation}
\times \rightarrow \lambda V_{\beta }=\lambda (K_{F})_{\mu \nu \alpha \beta
}\sigma ^{\mu \nu }q^{\alpha },  \label{V1}
\end{equation}%
in the momentum space.

We are interested in analyzing how the electron-positron scattering, $\
e^{+}+e^{-}\rightarrow\mu^{+}+\mu^{-}$, is altered by this new vertex. This
process may be depicted by the following tree-level Feynman diagrams:
\begin{figure}[H]
\begin{center}
\includegraphics[width=8cm]{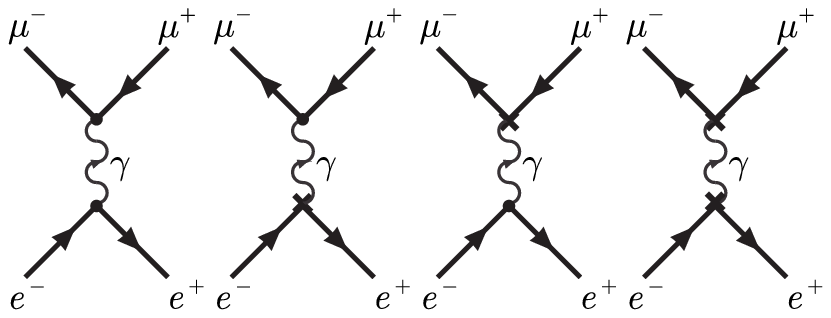}
\end{center}
\end{figure}

The tensor $K_{F}$\ is composed of birefringent and nonbirefringent
components. Without loss of generality, we restrain our investigation to the
nonbirefringent sector \cite{Obs1}, represented by nine coefficients and
parametrized by a symmetric and traceless rank-2 tensor defined by the
contraction
\begin{equation}
\kappa ^{\mu \nu }=\left( K_{F}\right) _{\alpha }{}^{\mu \alpha \nu },
\label{k2}
\end{equation}%
which fulfills
\begin{equation}
\left( K_{F}\right) ^{\lambda \nu \delta \rho }=\frac{1}{2}\left[ g^{\lambda
\delta }\kappa ^{\nu \rho }-g^{\nu \delta }\kappa ^{\lambda \rho }+g^{\nu
\rho }\kappa ^{\lambda \delta }-g^{\lambda \rho }\kappa ^{\nu \delta }\right]
.  \label{Alt}
\end{equation}%
Hence, the interaction (\ref{V1}) is rewritten as
\begin{equation}
{{\mathcal{L}}}_{I}^{new}=\lambda \kappa _{\nu \beta }\bar{\psi}\sigma ^{\mu
\nu }\psi F_{\mu }{}^{\beta },  \label{LI2}
\end{equation}%
which implies the following vertex:
\begin{equation}
\lambda V^{\mu }=\lambda q^{\beta }\left( \kappa ^{\nu \mu }\sigma _{\beta
\nu }-\kappa _{\nu \beta }\sigma ^{\mu \nu }\right) .  \label{V2}
\end{equation}

The components of the tensor can be classified by their parity properties: $%
\kappa_{00},\kappa_{ij}$ are parity even, while $\kappa_{0i}$ is parity odd.

\section{The cross section evaluation}

In this section, we evaluate the differential and total cross section for
the process,
\begin{equation}
e^{+}+e^{-}\rightarrow \mu ^{+}+\mu ^{-},
\end{equation}%
where the particles are labeled with momentum and spin variables as $%
e^{+}\left( p_{1};s_{1}\right) $, $e^{-}\left( p_{2};s_{2}\right) ,~\mu
^{+}\left( p_{1}^{\prime };s_{1}^{\prime }\right) $, and $\mu ^{-}\left(
p_{2}^{\prime };s_{2}^{\prime }\right) $. We work in the center of mass
frame, in which it holds $p_{1}=\left( E,\mathbf{p}\right) $, $p_{2}=\left(
E,-\mathbf{p}\right) $, $p_{1}^{\prime }=\left( E,\mathbf{p}^{\prime
}\right) $, and $p_{2}^{\prime }=\left( E,-\mathbf{p}^{\prime }\right) $,
with $p_{1},p_{2}$, and $p_{1}^{\prime },p_{2}^{\prime }$ being the momenta
of the incoming and outgoing particles, respectively. Transfer momentum $%
\left( q=p_{1}+p_{2}\right) $ is $q^{\beta }=\left( \sqrt{s},0\right) ,$
where $\sqrt{s}$ is the energy in the center of mass. In this frame, it
holds $\left\vert \mathbf{p}^{\prime }\right\vert ^{2}=\left\vert \mathbf{p}%
\right\vert ^{2}-m_{\mu }^{2}+m_{e}^{2},$ and
\begin{equation}
\rlap{$/$}p_{2}=\gamma ^{0}\rlap{$/$}p_{1}\gamma ^{0}=\gamma ^{0}\rlap{$/$}%
p\gamma ^{0},\text{ \ }\rlap{$/$}p_{2}^{\prime }=\gamma ^{0}\rlap{$/$}%
p_{1}^{\prime }\gamma ^{0}=\gamma ^{0}\rlap{$/$}p^{\prime 0},  \label{pp}
\end{equation}%
with ${m_{\mu }},m_{e}$ being the masses of the muon and the electron,
respectively. The vertex components are $V^{0}=0,$ and
\begin{equation}
V^{i}=\sqrt{s}\left( \kappa _{00}\sigma ^{0i}-\kappa _{ij}\sigma
^{0j}-\kappa _{0j}\sigma ^{ij}\right) .  \label{V3}
\end{equation}

Note that it holds $\kappa^{00}=\kappa^{ii}=\frac{3}{2}\kappa_{\text{tr}}, $
$\kappa^{ij}=-\left( \kappa_{e-}\right) ^{ij}+\frac{1}{2}\kappa_{\text{tr}%
}\delta^{ij}$, $\kappa^{0i}=-\kappa^{i},$ where $\kappa_{\text{tr}}$, and $%
\left( \kappa_{e-}\right) _{ij}$ correspond to the isotropic and anisotropic
parity-even components of the \emph{CPT}-even sector, respectively, while $%
\kappa^{i}$ represents the parity-odd components in accordance with Ref.
\cite{KM1}. These vertex components can be read as
\begin{equation}
V^{i}=V_{+I}^{i}+V_{+A}^{i}+V_{-}^{i},  \label{V5}
\end{equation}
where $V_{+I}^{i}=\sqrt{s}\kappa_{00}\sigma^{0i}$\ is the part associated
with the parity-even isotropic coefficient, $V_{+A}^{i}=-\sqrt{s}%
\kappa_{ij}\sigma^{0j}$ is related to the anisotropic parity-even component,
and $V_{-}^{i}=-\sqrt{s}\kappa_{0j}\sigma^{ij}=\sqrt{s}\kappa_{j}\sigma^{ij}$
is the contribution stemming from the parity-odd components.

In this scenario, the differential cross section (in natural units) is given
by
\begin{equation}
\frac{d\sigma }{d\Omega }=\frac{\left\vert \mathbf{p}^{\prime }\right\vert }{%
\left( 8\pi \right) ^{2}s\left\vert \mathbf{p}\right\vert }\left\vert {{%
\mathcal{M}}}\right\vert ^{2}.  \label{CS1}
\end{equation}%
The scattering amplitude is read off from the Feynman diagrams,
\begin{equation}
{{\mathcal{M}}}=\underset{a,b}{{\sum }}[\bar{v}^{s_{2}}(p_{2})\Gamma
_{\left( a\right) }^{\mu }u^{s_{1}}(p_{1})]\frac{1}{q^{2}}[\bar{u}%
^{s_{1}^{\prime }}(p_{1}^{\prime })\Gamma _{\left( b\right) \mu }v
^{s_{2}^{\prime }}(p_{2}^{\prime })],  \label{amplitudes}
\end{equation}%
where $a,b=0,1$ and $\Gamma _{\left( a\right) }^{\mu }$ defined by
\begin{equation}
\Gamma _{(0)}^{\mu }=-ie\gamma ^{\mu },\text{ }\Gamma _{(1)}^{\mu }=\lambda
V^{\mu },  \label{Vertices}
\end{equation}%
stands for the usual and new vertices. Here, $u^{s_{1}}(p_{1})$, $\bar{v}%
^{s_{2}}(p_{2})$ are the spinors for the electron and the positron, while $%
\bar{u}^{s_{1}^{\prime }}(p_{1}^{\prime })$, $v ^{s_{2}^{\prime }}\left(
p_{2}^{\prime }\right) $ represent the muon and antimuon spinors. For
evaluating the unpolarized cross section , the relevant quantity is $\langle
|{{\mathcal{M}}}|^{2}\rangle $, defined as $\left\vert {{\mathcal{M}}}%
\right\vert ^{2}={\displaystyle\sum }{{\mathcal{M}}{\mathcal{M}}}^{\ast }$,
where the sum is over the spin indices, $s_{1},s_{2},s_{1}^{\prime
},s_{2}^{\prime }$. This squared amplitude is carried out by means the
Casimir's trick, based on the use of spinor completeness relations and the
trace properties of $\gamma $ matrices. Knowing that
\begin{equation}
{{\mathcal{M}}}^{\ast }=\sum [\bar{u}^{s_{1}}(p_{1})\bar{\Gamma}_{\left(
a\right) }^{\mu }v^{s_{2}}(p_{2})]\frac{1}{q^{2}}[\bar{v}^{s_{2}^{\prime
}}(p_{2}^{\prime })\bar{\Gamma}_{\left( b\right) \mu }u^{s_{1}^{\prime
}}(p_{1}^{\prime })],  \label{amplitudeC}
\end{equation}%
the squared amplitude is written as
\begin{align}
\langle |{{\mathcal{M}}}|^{2}\rangle & =\frac{1}{4q^{4}}\sum \bar{v}%
^{s_{2}}(p_{2})\Gamma _{(a)}^{\mu }u^{s_{1}}(p_{1})\bar{u}^{s_{1}}(p_{1})%
\bar{\Gamma}_{(b)}^{\rho }v^{s_{2}}(p_{2})  \notag \\
& \times \bar{u}^{s_{1}^{\prime }}(p_{1}^{\prime })\Gamma _{(c)\mu
}v^{s_{2}^{\prime }}(p_{2}^{\prime })\bar{v}^{s_{2}^{\prime }}(p_{2}^{\prime
})\bar{\Gamma}_{(d)\rho }u^{s_{1}^{\prime }}(p_{1}^{\prime }).
\label{Msquared1}
\end{align}%
where $\bar{\Gamma}_{(i)}^{\mu }=\gamma ^{0}\Gamma _{(i)}^{\mu \dag }\gamma
^{0},$ and the sum is over the spin indices and over $a,b,c,d$. Using the
relation,
\begin{align}
& \bar{v}^{s_{2}}(p_{2})\Gamma _{(a)}^{\mu }u^{s_{1}}(p_{1})\bar{u}%
^{s_{1}}(p_{1})\bar{\Gamma}_{(b)}^{\nu }v^{s_{2}}(p_{2})  \notag \\
& =\text{tr}(\Gamma _{(a)}^{\mu }u^{s_{1}}(p_{1})\bar{u}^{s_{1}}(p_{1})\bar{%
\Gamma}_{(b)}^{\nu }v^{s_{2}}(p_{2})\bar{v}^{s_{2}}(p_{2})),
\end{align}%
the spin sum yields
\begin{equation}
\langle |{{\mathcal{M}}}|^{2}\rangle =\frac{1}{4q^{4}}L_{T}^{\mu \nu }\left(
M_{T}\right) _{\mu \nu },  \label{amplitudespins}
\end{equation}%
\begin{align}
L_{T}^{\mu \nu }& =L_{(00)}^{\mu \nu }+L_{(01)}^{\mu \nu }+L_{(10)}^{\mu \nu
}+L_{(11)}^{\mu \nu }, \\
M_{T}^{\mu \nu }& =M_{(00)}^{\mu \nu }+M_{(01)}^{\mu \nu }+M_{(10)}^{\mu \nu
}+M_{(11)}^{\mu \nu },
\end{align}%
with
\begin{align}
& \left. {L_{(ab)}^{\mu \nu }}{=}\text{{tr}}{[\Gamma _{(a)}^{\mu }(\rlap{$/$}%
p_{1}+m_{e})\bar{\Gamma}_{(b)}^{\nu }(\rlap{$/$}p_{2}-m_{e})],}\right.
\label{L1ab} \\
& {\;}\left. {M_{(ab)\mu \nu }=}\text{{tr}}{[\Gamma _{(a)\mu }(\rlap{$/$}%
p_{2}^{\prime }-m_{\mu })\bar{\Gamma}_{(b)\nu }(\rlap{$/$}p_{1}^{\prime
}+m_{\mu })].}\right.  \label{M1ab}
\end{align}%
Remember that the Latin indices inside parentheses $(a,b)$ can assume only
two values, $0$ or $1$, corresponding to the usual and new nonminimal
vertex, properly defined in Eqs. (\ref{V3}) and (\ref{Vertices}).

Next, in order to facilitate our evaluations and better discuss our results,
we proceed to separate the contributions coming from the parity-odd and
parity-even coefficients.

\subsection{Parity-odd contribution}

To calculate parity-odd contributions to the cross section, we restrict the
vertex (\ref{V3}) to
\begin{equation}
V_{-}^{i}=\sqrt{s}\sigma ^{ij}\kappa _{j},  \label{V6}
\end{equation}%
where $\kappa ^{i}=\left( K_{F}\right) ^{0jij}$. Using the trace technique,
and using identity (\ref{pp}), we show that%
\begin{equation}
L_{(01)}^{ij}=L_{(10)}^{ij}=M_{(01)}^{ij}=M_{(10)}^{ij}=0.  \label{L011}
\end{equation}%
The nonnull terms of the tensors (\ref{L1ab},\ref{M1ab}) are
\begin{align}
& \left. {L_{(00)}^{ij}}{=e}^{2}\text{{tr}}{[\gamma ^{i}\rlap{$/$}%
p_{1}\gamma ^{j}\rlap{$/$}p_{2}-m_{e}^{2}\gamma ^{i}\gamma ^{j}],}\right.  \\
& \left. {L_{(11)}^{ij}=\lambda }^{2}\text{{tr}}{[V^{i}\rlap{$/$}p_{1}V^{j}%
\rlap{$/$}p_{2}-m_{e}^{2}V^{i}V^{j}],}\right.  \\
& \left. {M_{(00)}^{ij}}{={e}^{2}}\text{{tr}}{[\gamma ^{i}\rlap{$/$}%
p_{1}^{\prime }\gamma ^{j}\rlap{$/$}p_{2}^{\prime }-m_{\mu }^{2}\gamma
^{i}\gamma ^{j}],}\right.  \\
& \left. {M_{(11)}^{ij}={\lambda }^{2}}\text{{tr}}{[V^{i}\rlap{$/$}%
p_{1}^{\prime }V^{j}\rlap{$/$}p_{2}^{\prime }-m_{\mu }^{2}V^{i}V^{j}]}%
,\right.
\end{align}%
while $L_{(ab)}^{0\mu }=L_{(ab)}^{\mu 0}=M_{(ab)}^{0\mu }=M_{(ab)}^{\mu 0}=0$%
. These latter terms are explicitly carried out:
\begin{align}
L_{(00)}^{ij}& ={e}^{2}\left( 2s\delta ^{ij}-8p^{i}p^{j}\right) ,  \notag \\%
[-0.3cm]
& \\
M_{(00)}^{ij}& ={e}^{2}\left( 2s\delta ^{ij}-8p^{\prime i}p^{\prime
j}\right) ,  \notag
\end{align}%
\begin{align}
L_{(11)}^{ij}& =8{\lambda }^{2}s\varepsilon ^{ikm}\varepsilon
^{jln}p^{n}p^{m}\kappa ^{k}\kappa ^{l},\text{ } \\
M_{(11)}^{ij}& =8{\lambda }^{2}s\varepsilon ^{ikm}\varepsilon
^{jln}p^{\prime n}p^{\prime m}\kappa ^{k}\kappa ^{l}.
\end{align}%
The squared amplitude is
\begin{align}
\langle |{{\mathcal{M}}}|^{2}\rangle & =\frac{1}{4s^{2}}\left[ L_{\left(
00\right) }^{ij}M_{\left( 00\right) }^{ij}+L_{\left( 11\right)
}^{ij}M_{\left( 00\right) }^{ij}\right.   \notag \\
& \left. +L_{\left( 00\right) }^{ij}M_{\left( 11\right) }^{ij}+L_{\left(
11\right) }^{ij}M_{\left( 11\right) }^{ij}\right] .  \label{Msquared2}
\end{align}

The differential cross section is obtained replacing these results in Eqs. ( %
\ref{Msquared2}) and (\ref{CS1}). The total cross section is obtained by
integration,
\begin{equation}
\sigma =\frac{\left\vert \mathbf{p}^{\prime }\right\vert }{\left( 8\pi
\right) ^{2}s\left\vert \mathbf{p}\right\vert }\int \left\langle \left\vert
\mathcal{M}\right\vert ^{2}\right\rangle d\Omega .
\end{equation}%
Taking the background as fixed, we integrate only on the angular variables
of the scattered particles, that is,%
\begin{align}
\int \left\langle \left\vert {{\mathcal{M}}}\right\vert ^{2}\right\rangle
d\Omega & =\frac{1}{4s^{2}}\left[ L_{00}^{ij}\int M_{00}^{ij}d\Omega
+L_{11}^{ij}\int M_{00}^{ij}d\Omega \right.  \notag \\
& +\left. L_{00}^{ij}\int M_{11}^{ij}d\Omega +L_{11}^{ij}\int
M_{11}^{ij}d\Omega \right] .
\end{align}%
These integrals provide%
\begin{align}
\int M_{\left( 00\right) }^{ij}d\Omega & =\frac{16{{e}^{2}}}{3}\left(
s+2m_{\mu }^{2}\right) \pi \delta ^{ij}, \\
\int M_{\left( 11\right) }^{ij}d\Omega & =\frac{8{{\lambda }^{2}}}{3}\pi
s\left( s-4m_{\mu }^{2}\right) \left( \delta ^{ij}\boldsymbol{\kappa }%
^{2}-\kappa ^{i}\kappa ^{j}\right) ,
\end{align}%
where the integral $\int p^{\prime i}p^{\prime j}d\Omega =\frac{1}{3}\left(
s-4m_{\mu }^{2}\right) \pi \delta ^{ij}$ was used. In the ultrarelativistic
limit, we take $m_{e}=m_{\mu }=0$. The resulting cross section (at second
order) is%
\begin{equation}
\sigma =\sigma _{QED}\left[ 1+\frac{1}{4e^{2}}\lambda ^{2}\left( 3s%
\boldsymbol{\kappa }^{2}-4\left( \mathbf{p}\cdot \boldsymbol{\kappa }\right)
^{2}\right) \right] .
\end{equation}

The results can be presented in two ways, concerning the beam orientation in
relation to the background vector, $\kappa^{i}.$ For the case where the beam
is perpendicular to the background, $\boldsymbol{\kappa}\cdot\mathbf{p}=0$,
we achieve
\begin{equation}
\sigma=\sigma_{QED}\left( 1+\frac{3s}{4e^{2}}\lambda^{2}|\boldsymbol{\kappa }%
|^{2}\right) ,  \label{CSortog}
\end{equation}
while for the case where the beam is parallel to the background, $%
\boldsymbol{\kappa }\cdot\mathbf{p}=\left\vert \boldsymbol{\kappa}%
\right\vert \sqrt{s}/2, $ the total cross section is
\begin{equation}
\sigma=\sigma_{QED}\left( 1+\frac{s}{2e^{2}}\lambda^{2}|\boldsymbol{\kappa }%
|^{2}\right) .  \label{CSlong}
\end{equation}

Experimental data from Ref. \cite{Derrick} for the $e^{+}+e^{-}\,\rightarrow
\mu^{+}+\,\mu^{-}$ scattering yields
\begin{equation}
\frac{\sigma-\sigma_{QED}}{\sigma_{QED}}=\pm\frac{2s}{\Lambda_{\pm}^{2}},
\label{Experiment}
\end{equation}
where $\sqrt{s}=29\; \mbox{GeV}$ and $\Lambda_{+}=170\; \mbox{GeV}$ with $%
95\%$ confidence level. Comparing (\ref{CSortog}) and (\ref{CSlong}) with (%
\ref{Experiment}), we obtain the following upper bound:
\begin{equation}
|\lambda\boldsymbol{\kappa}|<3\times10^{-12}\,(\mbox{eV})^{-1}.
\label{Bound1}
\end{equation}

\subsection{Parity-even contribution}

We begin considering the parity-even and isotropic contribution, whose
associated vertex is $V_{+I}^{i}=\sqrt{s}\kappa _{00}\sigma ^{0i}.$ In this
case, the elements of the tensors (\ref{L1ab},\ref{M1ab}) are
\begin{align}
L_{(00)}^{\mu \nu }& =e^{2}\text{tr}[\gamma ^{\mu }\rlap{$/$}p_{1}\gamma
^{\nu }\rlap{$/$}p_{2}-m_{e}^{2}\gamma ^{\mu }\gamma ^{\nu }], \\
L_{(01)}^{\mu \nu }& =ie\lambda m_{e}\text{tr}[\gamma ^{\mu }\rlap{$/$}%
p_{1}V_{+I}^{\nu }-\gamma ^{\mu }V_{+I}^{\nu }\rlap{$/$}p_{2}], \\
L_{(10)}^{\mu \nu }& =-ie\lambda m_{e}\text{tr}[V_{+I}^{\mu }\rlap{$/$}%
p_{1}\gamma ^{\nu }-V_{+I}^{\mu }\gamma ^{\nu }\rlap{$/$}p_{2}], \\
L_{(11)}^{\mu \nu }& =\lambda ^{2}\text{tr}[V_{+I}^{\mu }\rlap{$/$}%
p_{1}V_{+I}^{\nu }\rlap{$/$}p_{2}-m_{e}^{2}V_{+I}^{\mu }V_{+I}^{\nu }],
\end{align}%
The components of tensor $M_{(ab)}^{\mu \nu }$ are written in the same way,
changing $p_{1},p_{2},m_{e}$ by $p_{1}^{\prime },p_{2}^{\prime },m_{\mu }.$
In this case,
\begin{equation}
L_{\left( ab\right) }^{0\mu }=L_{\left( ab\right) }^{\mu 0}=M_{\left(
ab\right) }^{0\mu }=M_{\left( ab\right) }^{\mu 0}=0,
\end{equation}%
remaining as nonnull only the components $L_{\left( ab\right) }^{ij}$, $%
M_{\left( ab\right) }^{ij}$, given as
\begin{align}
L_{(01)}^{ij}& =L_{(01)}^{ij}=4e\lambda \kappa _{00}sm_{e}\delta ^{ij}, \\
M_{(01)}^{ij}& =M_{(01)}^{ij}=4e\lambda \kappa _{00}sm_{\mu }\delta ^{ij}, \\
L_{(11)}^{ij}& =8s\lambda ^{2}\left( \kappa _{00}\right) ^{2}\left(
m_{e}^{2}\delta ^{ij}+p^{i}p^{j}\right) , \\
M_{(11)}^{ij}& =8s\lambda ^{2}\left( \kappa _{00}\right) ^{2}\left( m_{\mu
}^{2}\delta ^{ij}+p^{\prime i}p^{\prime j}\right) .
\end{align}

The squared amplitude now is
\begin{align}
\langle| {{\mathcal{M}}}|^{2}\rangle & =\frac{1}{4s^{2}}\left(
L_{(00)}^{ij}+2L_{(01)}^{ij}+L_{(11)}^{ij}\right) \   \notag \\
& \times\left( M_{(00)}^{ij}+2M_{(01)}^{ij}+M_{(11)}^{ij}\right) .
\end{align}

Proceeding with the integration evaluations, and taking the
ultrarelativistic limit $\left( m_{e}=m_{\mu }=0\right) ,$ the total cross
section (at second order) is%
\begin{equation}
\sigma =\sigma _{QED}\left( 1+\frac{s}{e^{2}}\left\vert \lambda \kappa
_{00}\right\vert ^{2}\right) .
\end{equation}%
By using the same conditions as in Eq. (\ref{Experiment}), we achieve
\begin{equation}
|\lambda \kappa _{00}|<2.5\times 10^{-12}(\mbox{eV})^{-1}.
\end{equation}

We continue regarding the anisotropic parity-even contribution, whose vertex
is $V_{-A}^{i}=-\sqrt{s}\kappa ^{ij}\sigma ^{0j}.$ In this case, for turning
feasible the evaluations, we consider the ultrarelativistic limit $\left( {%
m_{e}}={m_{\mu }}=0\right) $ well in the beginning. The operators (\ref{L1ab}%
,\ref{M1ab}) are rewritten as
\begin{align}
L_{(00)}^{\mu \nu }& \approx e^{2}\text{tr}[\gamma ^{\mu }\rlap{$/$}%
p_{1}\gamma ^{\nu }\rlap{$/$}p_{2}], \\
L_{(11)}^{\mu \nu }& \approx \lambda ^{2}\text{tr}[V_{-A}^{\mu }\rlap{$/$}%
p_{1}V_{-A}^{\nu }\rlap{$/$}p_{2}],
\end{align}%
with components of the tensor $M_{(ab)}^{\mu \nu }$ read similarly by
changing $p_{1},p_{2},m$ by $p_{1}^{\prime },p_{2}^{\prime },M.$ Some
evaluations lead to
\begin{align}
L_{\left( ab\right) }^{0\mu }& =L_{\left( ab\right) }^{\mu 0}=M_{\left(
ab\right) }^{0\mu }=M_{\left( ab\right) }^{\mu 0}=0, \\
L_{(11)}^{ij}& =8s\lambda ^{2}\kappa ^{ik}\kappa ^{jl}p^{l}p^{k}, \\
M_{(11)}^{ij}& =8s\lambda ^{2}\kappa ^{ik}\kappa ^{jl}p^{\prime l}p^{\prime
k},
\end{align}%
implying
\begin{align}
& \left. \langle |{{\mathcal{M}}}|^{2}\rangle =\frac{1}{4s^{2}}\left[
L_{00}^{ij}M_{00}^{ij}\right. \right.  \notag \\
& \left. +8s\left( \kappa \right) ^{ik}\left( \kappa \right) ^{jl}\left(
p^{l}p^{k}M_{00}^{ij}+L_{00}^{ij}p^{\prime l}p^{\prime k}\right) \right] .
\end{align}

Doing the corresponding integrations in the solid angle, we achieve
\begin{equation}
\int \!\!\left\langle \left\vert {{\mathcal{M}}}\right\vert
^{2}\right\rangle d\Omega =\frac{16\pi e^{4}}{3}\left[ 1+\frac{\lambda ^{2}}{%
4e^{2}}\left( s\left( \kappa ^{2}\right) ^{ii}+4\left( \kappa
^{ij}p^{j}\right) ^{2}\right) \right] ,
\end{equation}%
where $\left( \kappa ^{2}\right) ^{ii}=\kappa ^{ij}\kappa ^{ji}$. Choosing a
beam direction so that $\kappa ^{ij}p_{j}=0,$ we attain
\begin{equation}
\int \left\langle \left\vert {{\mathcal{M}}}\right\vert ^{2}\right\rangle
d\Omega =\frac{16e^{4}\pi }{3}\left( 1+\frac{\lambda ^{2}s}{4e^{2}}\left(
\kappa ^{2}\right) ^{ii}\right) .  \label{4.140)}
\end{equation}%
$\ $This evaluation leads to
\begin{equation}
\sigma =\sigma _{QED}\left( 1+\frac{\lambda ^{2}s}{4e^{2}}\left( \kappa
^{2}\right) ^{ii}\right) ,  \label{4.142)}
\end{equation}%
implying the following upper bound:
\begin{equation}
|\lambda \kappa ^{ij}|<5\times 10^{-12}(\mbox{eV})^{-1}.  \label{4.143)}
\end{equation}

We notice that the upper bound on the parity-even parameters have the same
order of magnitude as the one on the parity-odd coefficients.

\section{Conclusions}

In this work, we have studied the influence of a Lorentz-violating \textit{%
CPT}-even nonminimal coupling in the context of the Dirac equation, focusing
specifically on the $e^{+}+e^{-}\rightarrow \mu ^{+}+\mu ^{-}$ scattering
process. This new coupling implied the insertion of a new vertex, increasing
the number of Feynman diagrams representing the level tree process. We have
carried out the contributions of the nonminimally \textit{CPT}-even LV terms
on the unpolarized cross section, using the Casimir's trick. This evaluation
was performed with details for the parity-odd and parity-even\ coefficients
in the ultrarelativistic limit $\left( {m_{e}}={m_{\mu }}=0\right) $.
Comparing the attained results with scattering data in the literature \cite%
{Derrick}, we have succeeded in imposing upper bounds at the level of $%
10^{-12}\,(\mbox{eV})^{-1}$ on the parity-odd and parity-even
nonbirefringent coefficients of the quantity $\lambda \left( K_{F}\right)
_{\mu \nu \alpha \beta },$ representing a\ good route to constrain the
strength of this new nonminimal coupling in a relativistic environment. It
is important to mention that these bounds should not be directly compared
with the upper bounds imposed on the coefficients of the dimensionless
\textit{CPT}-even tensor $\left( K_{F}\right) _{\mu \nu \alpha \beta }$\ in
Refs. \cite{KM1,Risse}. The bounds here achieved restrain the dimensional
quantity $\lambda \left( K_{F}\right) _{\mu \nu \alpha \beta }$,
representing a constraint on the way the CPT-even is coupled to the fermion
sector.

\ Although we have restricted our study to the nonbirefringent sector of the
\textit{CPT}-even tensor $\left( K_{F}\right) _{\mu \nu \alpha \beta },$\ we
could have considered the ten birefringent components of the tensor $\left(
K_{F}\right) _{\mu \nu \alpha \beta }$\ as well. The point is that these
coefficients contribute to the modified cross section also in second order,
implying the same upper bound attained on the nonbirefringent components.
This reasoning allows to extend the bounds here achieved to all the
components of the tensor $\left( K_{F}\right) ,$ that is $\left\vert \lambda
\left( K_{F}\right) \right\vert \leq 10^{-12}\left( \mbox{eV}\right) ^{-1},$
circumventing some cumbersome and unnecessary evaluations.

An interesting investigation concerns the possible connections between this
dimension-5 nonminimal coupling and the higher-dimensional operators
belonging to the photon sector presented in Ref. \cite{Kostelec}. The
proposed nonminimal coupling is a dimension-5 operator which is not
contained in the framework of Refs. \cite{Kostelec}, once this term refers
to the interaction between fermions and photons. The connection begins to
appear when one performs the radiative corrections generated by this
nonminimal coupling. Indeed, the evaluation of the one-loop vacuum
polarization diagram of the photon leads to operators with dimension-4 and
-6. The dimension-4 operator is exactly the \textit{CPT}-even term $%
(K_{F})_{\mu \nu \rho \sigma }F^{\mu \nu }F^{\rho \sigma }$. The operators
of dimension-six are second order in $K_{F}$\ and could be encompassed in
Ref. \cite{Kostelec}. The fact that the dimension-4 operator can be
generated by radiative corrections allows one to use the existing bounds on
the \textit{CPT}-even $\left( K_{F}\right) _{\mu \nu \alpha \beta }$ to
attain even better bounds on the magnitude of the quantity $\lambda \left(
K_{F}\right) _{\mu \nu \alpha \beta }.$ The detailed analysis of this issue
is under development now \cite{Frede2}.

\begin{acknowledgments}
The authors are grateful to CNPq, CAPES and FAPEMA (Brazilian research
agencies) for invaluable financial support.{}
\end{acknowledgments}


\begin{thebibliography}{99}
\bibitem{Obs1} An additional reason for limiting our evaluations on the
nonbirefringent sector is provided in the conclusions.
\end{thebibliography}
\end{document}